\documentclass[11pt,onecolumn,amssymb,nofootinbib]{revtex4}
\usepackage{amsmath, amsthm, amscd, amssymb}
\usepackage{bm}
\usepackage{bbm}

\begin{document}

\title{\bf Phases with modular ground states for symmetry breaking by  rank 3 and rank 2  antisymmetric tensor scalars }

\author{Stephen L. Adler}
\email{adler@ias.edu} \affiliation{Institute for Advanced Study,
Einstein Drive, Princeton, NJ 08540, USA.}

\begin{abstract}

Working with explicit examples given by the 56 representation in $SU(8)$, and the 10 representation in $SU(5)$, we show that  symmetry breaking of a group ${\cal G}\supset {\cal G}_1 \times {\cal G}_2$ by a scalar in a rank three or two antisymmetric tensor
representation leads to a number of distinct   {\it modular} ground states.  For these broken symmetry phases,  the ground state is  periodic in an integer divisor $p$ of $N$, where $N>0$ is the absolute value of the nonzero $U(1)$ generator of the scalar component $\Phi$  that is a singlet under the simple subgroups ${\cal G}_1$ and ${\cal G}_2$.  Ground state expectations of fractional powers $\Phi^{p/N}$ provide order parameters that
distinguish the different phases.  For the  case of period $p=1$, this reduces to the
usual Higgs mechanism, but for  divisors $N\geq p>1$ of $N$ it leads to a modular ground state with periodicity $p$, implementing a discrete Abelian symmetry group $U(1)/Z_p$.   This observation may  allow new approaches to grand unification and family unification.
\end{abstract}

\maketitle

\section{Introduction}

The possibilities for constructing grand unified models depend crucially on the pathways available for symmetry breaking, and the
corresponding structures of the vacuum or ground state.  For the familiar case of $SU(5)$ grand unification, breaking to the standard
model is accomplished by assuming a scalar in the 24 representation, which has a singlet component under the $SU(2) \times SU(3)$ subgroup,
with zero $U(1)$ generator.  Hence the group after symmetry breaking, when the singlet scalar component acquires a
nonzero vacuum expectation in a vacuum in a $U(1)$ eigenstate with eigenvalue 0,  is $SU(2)\times SU(3) \times U(1)$.  Discussions
of symmetry breaking by second rank \cite{li}  and third rank \cite{cummins} antisymmetric tensors have  assumed
a ground state that completely breaks the analogous $U(1)$, since the singlet component of the scalar
under the simple subgroups of the initial gauge group typically has a {\it nonzero} $U(1)$ generator, with absolute value that
we denote by $N$. We shall show in this paper that the
phase with completely broken $U(1)$
is only the simplest of a set of symmetry breaking phases, the rest of which have discrete $U(1)/Z_p$ residual symmetry, with $p$ an integer divisor
of $N$,  corresponding to  a {\it modular} ground state that is periodic in the $U(1)$ generator with period $p$.

We begin in Sec. 2 by considering a model that we recently proposed \cite{adler1} for $SU(8)$ family unification, with $SU(8)$ broken
by a scalar in the rank three antisymmetric tensor 56 representation. We
review the arguments that consistency of symmetry breaking by a third rank antisymmetric tensor scalar field in the
56 representation, together with the requirement of clustering, requires a ground state structure of modularity 15 in the $U(1)$ generator.  This
can be achieved by a ground state that has periodicity $p$ in the $U(1)$ eigenvalue, with $p$ any integer divisor of 15, that is $p=1,\,3,\,5,\,15$.
We show that the case $p=1$ corresponds to the  calculation of  \cite{cummins}.  Transforming from $U(1)$ eigenvalue
space to $\omega$ space, where $-15 \omega$ is the phase angle of the component of the 56 that attains a vacuum expectation value, we see that this case
corresponds to the usual Higgs mechanism where the vacuum picks a value of $\omega$ in the range $0 \leq \omega < 2 \pi$.  The cases
of $p > 1$ correspond in $\omega$ space to the vacuum picking a value of $\omega$ in the range $0 \leq \omega < 2 \pi/p$, since the
period $p$ modularity of the ground state leads to a discrete Abelian symmetry $U(1)/Z_p$, which maps the wedge sectors $2 \pi n /p \leq \omega < 2 \pi (n+1) /p$,
for $n=0,1,...,p-1$, into one another.  This scenario has been analyzed in the context of Abelian models, for the case when $p$ is equal to the $U(1)$ generator of a condensing scalar field, in the presence of a second scalar field with $U(1)$ generator 1,  by Krauss and Wilczek \cite{krauss}, by
Banks \cite{banks}, and by Preskill and Krauss \cite{preskill}.  We review their conclusions showing that although the $U(1)$ gauge field gets a mass, the modularity of the ground state leads to
conservation of $U(1)$ generator charges modulo $p$, and to other residual long range effects.  Thus, the $p>1$ symmetry breaking phases contain
new physics not evident from the $p=1$ case.

In Sec. 3 we present analogous considerations for the case of $SU(5)$ broken by a rank two antisymmetric scalar in the 10 representation, and generalize to the case of
$SU(n)$ broken by a rank three or two antisymmetric tensor.  In Sec. 4 we give a brief discussion of implications of these results for model building, and in particular for the $SU(8)$ model proposed in \cite{adler1}. Finally, in an Appendix
we  give a finite sum construction of a modulo $p$ basis from an initially modulo $N$ basis, with $p$ a divisor of $N$, and show that this construction inherits the clustering properties of the modulo $N$ basis.

\section{SU(8) broken by a rank three antisymmetric tensor: modulo $p$ structure of the broken symmetry ground state, where $p\, |\, 15$ }

We briefly recall those elements of \cite{adler1} needed for the discussion here.   We start from an
$SU(8)$ gauge theory, with a gauge boson $A_{\mu}^A$, $A=1,...,63$, and a complex scalar field $\phi^{[\alpha\beta\gamma]}$,
$\alpha,\,\beta,\,\gamma =1,...,8$ in the totally antisymmetric 56 representation.  (There are also gauged fermion
fields in the model, but the details are not needed for our analysis here.)  We are interested in the symmetry
breaking pattern $SU(8) \supset SU(3) \times SU(5) \times U(1)$, under which the branching behaviors of the
63 and 56 $SU(8)$ representations are \cite{slansky},
\begin{align}\label{eq:branching}
 A_{\mu}^A:~~~&~~63=(1,1)(0)+(8,1)(0)+(1,24)(0)+(3,\overline{5})(-8)+(\overline{3},5)(8)~~~,\cr
\phi^{[\alpha\beta\gamma]}:~~~&~~56=(1,1)(-15)+(1,\overline{10})(9)+(\overline{3},5)(-7)+(3,10)(1)~~~.\cr
\end{align}
Here the numbers $(g)$ in parentheses following the representation labels $(m,n)$ are the values of the
$U(1)$ generator $G$ defined by the commutator
\begin{equation}\label{eq:gdef}
[G,(m,n)]= g\,(m,n)~~~.
\end{equation}
This commutator is  linear in the operator in representation $(m,n)$, and so is independent of its
normalization, but the values of $g$ depend on the normalization of the $U(1)$  generator $G$.
The $U(1)$ generator values in the branching Table 54 of Slansky \cite{slansky} are based on the $8\times 8$ matrix $U(1)$ generator
\begin{equation}\label{eq:norm}
G={\rm Diag} (-5,-5,-5,3,3,3,3,3)~~~,
\end{equation}
as can be read off from the branching rule for the fundamental 8 representation.

With the Slansky normalization, all $U(1)$
eigenvalues in the branching table are integers, attaining values $0, \pm 1, \pm 2, ...$.  Note that if we changed the $U(1)$
generator normalization by a factor  of $q$  (not necessarily an integer), the corresponding eigenvalues would be $0, \pm q , \pm 2q, ... $, and so the
spectrum in units $q$ would have the same structure as in the case $q=1$.  In particular, in terms of the 56 representation
analogs of the two fields of Refs.  \cite{krauss}, \cite{banks}, and \cite{preskill}, the ratio of the $U(1)$ generator $-15q$ of the $(1,1)$ to the $U(1)$ generator $q$ of the $(3,10)$ is still $-15$, independent of $q$.  All of the results that we obtain in this paper depend only on {\it relative} $U(1)$ generator values and are
independent of the normalization $q$; when $q\neq 1$, our statements about periodicity modulo $p$  with $p$ a divisor of 15, become
statements about periodicity modulo $pq$, with $p$ still an integer divisor of 15.  So noting this, we will always use the Slanksy normalization
$q=1$. This directly exhibits the role of $G$ as the phase angle rotation generator, with angles measured in units of radians, which would not be
the case for a normalization equating the trace of the square of the $U(1)$ generator to $\frac{1}{2}$, coresponding to an irrational, nonintegral value of $q$.

As pointed out in \cite{adler1}, the $\big(SU(3),SU(5)\big)$ singlet component $(1,1)$ of $\phi$, which we denote by $\Phi$,  has a $U(1)$ generator
value of $-15$  (corresponding to $N=15$), for which Eq. \eqref{eq:gdef} becomes
\begin{equation}\label{eq:gdef1}
[G,\Phi]=-15 \Phi.
\end{equation}
Writing
\begin{equation}\label{eq:absphidef}
\Phi=|\Phi|e^{-15i\omega}~~~,
\end{equation}
with real $\omega$, this corresponds to the $U(1)$ generator $G$ acting as
$G\sim -i \partial/\partial \omega$.  Equation \eqref{eq:gdef1} implies that
$\Phi$  cannot have a nonzero expectation in a state $|g^{\prime}\rangle$ with a definite $U(1)$ generator
value, since taking the expectation in this state gives
\begin{equation}\label{eq:problem}
0=\langle g^{\prime}| [G,\Phi] |g^\prime \rangle =  -15 \langle g^{\prime}| \Phi |g^\prime \rangle~~~.
\end{equation}
For symmetry to be broken, the ground state must be a superposition of states containing at least two $U(1)$
generator values differing by  $15$.  As shown in Appendix A of \cite{adler1} (which focused on the case $p=5$; we generalize here and use a
different notation), this requirement, plus a clustering argument similar to the one leading to the periodic
theta vacuum in quantum chromodynamics, dictates that
the ground state must be  an infinite sum of the form
\begin{equation}\label{eq:ground1}
|0,\omega\rangle_p= \sum_{n=-\infty}^{\infty}\left(\frac{p}{2\pi}\right)^{1/2} e^{ipn\omega}|pn\rangle ~~~,
\end{equation}
with $p$ an integer divisor of 15.
The corresponding basis of states is then, for $k=0,...,p-1$,
\begin{equation}\label{eq:basis1}
|k,\omega\rangle_p=\sum_{n=-\infty}^{\infty}\left(\frac{p}{2\pi}\right)^{1/2} e^{i(k+pn)\omega}|k+pn\rangle ~~~.
\end{equation}
Apart from the overall normalization, the terms in the sum of Eq. \eqref{eq:ground1} for $p>1$ are simply a subset of those
in this sum for $p=1$, and for general $p$ the action of  the $U(1)$ generator $G$ on the basis is that of a generator of rotations in
$\omega$,
\begin{align}\label{eq:genrotate}
G|k,\omega\rangle_p=&G\sum_{n=-\infty}^{\infty}\left(\frac{p}{2\pi}\right)^{1/2} e^{i(k+pn)\omega}|k+pn\rangle\cr
=&\sum_{n=-\infty}^{\infty}\left(\frac{p}{2\pi}\right)^{1/2} e^{i(k+pn)\omega}(k+pn)|k+pn\rangle
= -i\, \frac{\partial}{\partial \omega} |k,\omega\rangle_p~~~.\cr
\end{align}
The state basis obeys the modulo $p$ periodicity
\begin{equation}\label{eq:period}
|k+ps,\omega\rangle_p=|k,\omega\rangle_p
\end{equation}
for any integer $s$, and up to an infinite proportionality constant,
obeys the clustering property
\begin{equation}\label{eq:cluster1}
|k_A+k_B,\omega\rangle_p \propto |k_A,\omega \rangle_p  |k_B,\omega \rangle_p
 \end{equation}
 for widely separated subsystems $A,\, B$.
 In the modular basis of Eqs. \eqref{eq:ground1} and \eqref{eq:basis1}, we can have a nonzero
expectation of $\Phi$.

As noted in passing in \cite{adler1}, there is a second argument leading to the conclusion that the ground
state must have modularity $p|N$.
In order for $SU(8)$ to break to $SU(3) \times SU(5)$, the gauge bosons in the $(3,\overline{5})(-8)$ and
$(\overline{3},5)(8)$ representations must become massive, by picking up longitudinal components from the
corresponding representations in the branching of the scalar $\phi$ and its complex conjugate.
However, the representation of $\phi$ corresponding to $(\overline{3},5)(8)$ in the branching shown in Eq. \eqref{eq:branching} is
$(\overline{3},5)(-7)$, which has the same $SU(3)$ and $SU(5)$ representation content, but {\it a $U(1)$
generator differing by 15}.  Thus  consistency of the Brout-Englert-Higgs-Guralnik-Hagen-Kibble symmetry breaking mechanism to give
the vector bosons a mass also requires a modulo $p$ state structure in the $U(1)$ generator values $g$, with $p$ a divisor of 15.

\subsection{The case $p=1$}

The simplest case to consider is $p=1$, for which we simplify the notation by dropping the subscript $p=1$ on the state vector, so that
$|0,\omega\rangle_1 \equiv  |0,\omega\rangle$.    When $p=1$, the sum in Eq. \eqref{eq:ground1} becomes
\begin{equation}\label{eq:p1case1}
|0,\omega\rangle= \sum_{n=-\infty}^{\infty}\left(\frac{p}{2\pi}\right)^{1/2} e^{in\omega}|n\rangle~~~,
\end{equation}
and this ground state is the only state in the state basis. The inverse of Eq. \eqref{eq:p1case1} is
\begin{equation}\label{eq:p1case2}
|n\rangle=\int_0^{2\pi} d\omega \left(\frac{p}{2\pi}\right)^{1/2} e^{-in\omega}|0,\omega\rangle~~~.
\end{equation}
Modularity with $p=1$ means that  all $U(1)$ charge states are
equivalent, that is, there is no conserved $U(1)$ charge.   The state of Eq. \eqref{eq:p1case1} is the angular state where the
scalar component $\Phi$ has the form of Eq. \eqref{eq:absphidef}. This can be seen from the facts that $|0,\omega\rangle$ has
periodicity $2\pi$ in the phase angle $\omega$, with $G$ acting as the generator of rotations in $\omega$, as well as from the from the completeness and orthonormality  relations
\begin{align}\label{eq:complete1}
\int_{0}^{2\pi} d\omega |0,\omega\rangle \langle 0,\omega|=&\sum_{n=-\infty}^{\infty} |n\rangle \langle n|=1~~~,\cr
\langle \omega^{\prime}|\omega \rangle=& \frac{1}{2\pi} \sum_{n=-\infty}^{\infty} e^{in(\omega-\omega^{\prime})}=\delta(\omega-\omega^{\prime})~~~,\,0\leq\omega-\omega^{\prime}<2\pi~~~.
\end{align}
When the Higgs potential $V(\phi)$ has a minimum at the nonzero magnitude $|\Phi|=\rho$, then we get the usual Higgs mechanism analyzed in
\cite{cummins}:  the $U(1)$ gauge field gets a mass proportional to $\rho^2$, with the state of Eq. \eqref{eq:p1case1} corresponding to a vacuum
where $\Phi$ has  phase angle $-15 \omega$.

\subsection{The cases $p>1$}

When $p>1$,  the basis functions in the expansion of Eqs. \eqref{eq:ground1}, \eqref{eq:basis1} are
 $\left(\frac{p}{2\pi}\right)^{1/2} e^{ipn\omega}$, which
are complete and orthonormal on the interval
\begin{equation}\label{eq:wedge1}
0\leq \omega \leq 2\pi/p~~~.
\end{equation}
So the states
of  Eqs. \eqref{eq:ground1}, \eqref{eq:basis1} are phase angle eigenstates defined on  this
interval, and  have the angular periodicity
\begin{equation}\label{eq:period1}
|k,\omega+2\pi m/p\rangle_p=e^{2\pi i m k/p} |k,\omega\rangle_p~~~.
\end{equation}
The completeness and orthnormality relations now read
\begin{align}\label{eq:complete2}
\int_{0}^{2\pi/p} d\omega |k,\omega\rangle_p\, {}_p\langle k,\omega|=&\sum_{n=-\infty}^{\infty} |k+pn\rangle \langle k+pn|\equiv1_{k,p}~~~,\cr
\sum_{k=0}^{p-1} \int_{0}^{2\pi/p} d\omega |k,\omega\rangle_p\, {}_p\langle k,\omega|=&\sum_{k=0}^{p-1}1_{k,p}=1~~~,\cr
{}_p\langle k^{\prime},\omega^{\prime}|k,\omega \rangle_p= \delta_{k\,k^{\prime}} e^{ik(\omega-\omega^{\prime})}\frac{p}{2\pi} \sum_{n=-\infty}^{\infty} e^{ipn(\omega-\omega^{\prime})}=&\delta_{k,k^{\prime}} \delta(\omega-\omega^{\prime})~~~,\,0\leq \omega-\omega^{\prime}<2\pi/p~~~,
\end{align}
with $1_{k,p}$ the projector on the subspace of states with $U(1)$ generator values equivalent to $k$ modulo $p$.

Under the transformation $\omega \to \omega+2\pi/p$, Eq. \eqref{eq:period1} shows that the state $|k,\omega\rangle$
 is multiplied by a $p$th root of unity  $e^{2\pi i k/p}$, characterizing the
residual $U(1)/Z_p$ symmetry analyzed (for $p=N$) in \cite{krauss}, \cite{banks}, and \cite{preskill}.
The Higgs potential $V(\phi)$ is now minimized over the wedge shaped domain $0\leq |\Phi|< \infty,\, 0\leq \omega \leq 2\pi/p$,
with periodic boundary conditions on the edges of the wedge. This domain is not a manifold, but rather an orbifold with a conical
singularity at the origin of $|\Phi|$.  Since the Higgs potential has no dependence on $\omega$, the potential minimum is the same
as in the $p=1$ case discussed above:  the minimum is at $|\Phi|=\rho$ and the $U(1)$ gauge boson receives a mass proportional to
\begin{equation}\label{eq:massprop}
[G,\Phi]^{\dagger} [G,\Phi] = (15)^2 \rho^2~~~,
\end{equation}
just as in the $p=1$ case. It is precisely because the potential minimum is insensitive to $\omega$ that different
symmetry breaking phases, corresponding to different discrete Abelian groups of multiplicative factors composed of the $p$th roots of unity, are possible.  The broken symmetry phases corresponding to different values of $p$ are energetically degenerate; it
is only at subsequent stages of symmetry breaking, beyond that produced by $V(\phi)$, that one expects one particular value of
$p$ to be singled out as the most energetically favored ground state.

\subsection{The case p=N}

However, as discussed in \cite{krauss}, \cite{banks}, and \cite{preskill}, the fact that the $U(1)$ gauge boson becomes
massive is not the end of the story; there are residual effects at low energy scales (as defined by the gauge boson mass)
that reflect the hidden  $U(1)/Z_p$ symmetry.  Specifically, Banks \cite{banks} constructs the low energy effective action
implicit in the Abelian model of Krauss and Wilczek \cite{krauss}, and shows that (with his $q$ our $N$, and with his $\phi$ a  charge one scalar) ``as a consequence of gauge invariance all
terms in the low-energy lagrangian will have a global $Z_q$ symmetry and all terms relevant at low energy (when $q \geq 5$) will have a global
symmetry under phase rotations of $\phi$''.  And Preskill and Krauss \cite{preskill} note that for the model of \cite{krauss} \big(with their $N$ our $N$, their $Q$ our $k$ of Eq. \eqref{eq:basis1},
and their $k$ our $m$ of Eq. \eqref{eq:period1}\big) ``...the charge modulo $N$ of a state is not screened by the Higgs condensate.  Thus, there is a nontrivial superselection rule.  The Hilbert space decomposes into $N$ sectors labeled by
the charge $Q$ mod $N$, with states of charge $Q$ transforming under $Z_N$ gauge transformations according to  $U(\frac{2\pi k}{N})|Q\rangle=e^{2\pi i k Q/N} |Q\rangle$. Each sector is preserved by the
gauge-invariant local observables.''

 What we have shown in the preceding sections is that these statements of Banks  and of  Preskill and Krauss about the model of Krauss and Wilczek generalize to divisors $p$ of $N$, corresponding to alternative phases of lower symmetry with ground states obeying clustering.  Additionally,
 we have shown that these statements find a natural  application in the breaking of the $SU(8)$ gauge group by a scalar in the 56 representation, as a consequence of the nonzero $U(1)$ generator of the  symmetry breaking $(1,1)(-15)$ component.

\subsection{Order parameters}

When a system has distinct phases, there are order parameters that differentiate among them. To find
order parameters appropriate to $SU(8)$ broken by a 56 representation scalar, we note that
$\Phi^{p^{\prime}/15}$  has $U(1)$ generator $-p^{\prime}$ when $p^{\prime}=1,\,3,\,5,\,15$ is a divisor
of $15$.  Hence the expectation of $\Phi^{p^{\prime}/15}$ in the ground state $|0,\omega\rangle_p$ will
vanish unless $p^{\prime}/p$ is a positive integer $r$, in which case the spacing of levels in
$|0,\omega\rangle_p$ includes matches to $p^{\prime}$,
\begin{align}\label{eq:order1}
{}_p\langle 0,\omega|\Phi^{p^{\prime}/15}|0,\omega\rangle_p=&\sum_{n=-\infty}^{\infty}
e^{ipr\omega} \langle p(n-r)|\Phi^{p^{\prime}/15}|pn\rangle ~~~,~p^{\prime}/p={\rm integer}~r \geq 1~~~,\cr
{}_p\langle 0,\omega|\Phi^{p^{\prime}/15}|0,\omega\rangle_p=&0~~~{\rm~otherwise}~~~.\cr
\end{align}
This gives order parameters that
can distinguish between the $p=1,\,3,\,5,\,15$ symmetry breaking phases.

\section{ Modulo $p|6$ structure of $SU(5)$ broken by a rank two antisymmetric tensor, and the general $SU(n)$ case}

We focused in Sec. II on the $SU(8)$ case because that is needed for the analysis of \cite{adler1}, but our results are
more general.  For example, in the symmetry breaking pattern $SU(5) \supset SU(2) \times SU(3) \times U(1)$ by a complex scalar in the rank two antisymmetric 10 representation,
the relevant branching behaviors of the adjoint 24 and the 10 representations are \cite{slansky}
\begin{align}\label{eq:su5case}
24=&(1,1)(0)+(3,1)(0)+(1,8)(0)+(2,3)(-5)+(2,\overline{3})(5)~~~,\cr
10=&(1,1)(6)+(1,\overline{3})(-4)+(2,3)(1)~~~,\cr
\end{align}
and the corresponding $U(1)$ generator is
\begin{equation}\label{eq:su5u1}
G={\rm Diag}(3,3,-2,-2,-2)~~~.
\end{equation}
Since the $(1,1)$ component of the 10 representation has $U(1)$ generator 6, the broken
symmetry ground state must have modularity $p$  in this generator, with $p$ a divisor of 6.  In order for the adjoint
component $(2,3)(-5)$ to absorb the scalar component $(2,3)(1)$ to obtain a mass, modularity
$p|6$ in the $U(1)$ generator is again needed.

For the case of general $SU(n)$, we have computed
the branching expressions and value of $N$ generalizing Eq. \eqref{eq:branching}, Eq. \eqref{eq:su5case} and additional branching
expressions in  \cite{slansky}; the detailed results, which agree with the ground state modularity pattern found in the special cases discussed here, will be reported elsewhere. In the generic case of $SU(n)$ breaking by a rank three antisymmetric tensor component
with $U(1)$ generator of absolute value $N$,  the breaking pattern \cite{cummins} $SU(n)\supset SU(3) \times SU(n-3)$ is 
extended to $SU(n)\supset SU(3) \times SU(n-3)\times U(1)/Z_p$, with $p$ a divisor of $N$.  In the generic case of $SU(n)$ breaking by a rank two antisymmetric
tensor component with $U(1)$ generator of absolute value $N$,  the breaking pattern \cite{li} $SU(n)\supset SU(2) \times SU(n-2)$ is extended to $SU(n)\supset SU(2) \times SU(n-2)\times U(1)/Z_p$, with $p$ a divisor of $N$. 
 For general $n$, as in the $SU(8)$  example, the ground state
expectation of $\Phi^{p^{\prime}/N}$ serves as an order parameter for distinguishing among the
different phases, according to the generalization of Eq. \eqref{eq:order1}.

\section{Discussion}

We have shown that symmetry breaking by scalars in rank two and rank three antisymmetric tensor representations
requires in general a modular state structure in the broken symmetry phase, a fact that extends  previous analyses of symmetry breaking patterns.  The consequences
of this periodic ground state structure in the $U(1)$ generator open new, potentially experimentally viable possibilities for family and grand unification.

With particular reference to the model of \cite{adler1}, which uses the breaking pattern $SU(8)\supset SU(3) \times SU(5) \times U(1)/Z_5$, and connects
this with flipped $SU(5)$ grand unification \cite{barr}, \cite{der}, we remark that:
\begin{enumerate}
\item    Breaking flipped $SU(5)$ to the standard model uses the  Higgs $p=1$ phase of symmetry breaking by a 10
representation scalar, so considerations of modularity do not enter.
\item   Since the gauge field associated with the $U(1)/Z_5$ factor in the
initial $SU(8)$ symmetry breaking  acquires a mass, an additional mechanism may be needed to get a massless $U(1)_Y$ gauge field (with $Y$ the weak hypercharge)
after symmetry breaking to the standard model.   The remark of Banks \cite{banks} that for $p\geq 5$ the residual discrete Abelian symmetry becomes
a full global $U(1)$ symmetry may be relevant here, since the model of \cite{adler1} employs $p=5$. A
global $U(1)$ indicates that  $G$ annihilates the low energy states; evaluating the left hand side
of Eq. \eqref{eq:massprop} in the residual low energy theory then suggests an effective mass of zero for
the $U(1)$ gauge boson.  That is, although the ``bare'' mass of the $U(1)$ gauge boson is nonzero at the
unification scale, as indicated by the right hand side of Eq. \eqref{eq:massprop}, the effective mass
may run with energy, giving a value of zero for the ``dressed'' mass at low energies.  A further examination of this scenario in relation to the model of \cite{adler1} will be undertaken elsewhere.
\item   Another issue still to
be decided is whether the residual long range effects associated with the $U(1)/Z_5$ factor can give the alignment of states  needed   to obtain the correct
quantum numbers of the standard model fermions.
\item   Discrete Abelian gauge symmetries can be
 studied using dual $BF$ models \cite{uranga}, and these may be helpful in further analyzing the model of \cite{adler1}.
\end{enumerate}

\section{Acknowledgements}

I wish to thank Juan Maldecena for  helpful conversations that led me to look again at \cite{krauss} -- \cite{preskill}, and alerted
me to the literature on $BF$ models.  I also wish to acknowledge the hospitality in September 2014 of the   CERN  Theory Division  during work on this paper.

\appendix

\section{State basis modulo $p$ built from a basis modulo $N$, when $p$ divides $N$}

In \cite{adler1}, we postulated that the state basis after $SU(8)$ breaking has a modulo 5 invariance in the
$U(1)$ generator values, and as noted, we used this invariance to make a connection with flipped $SU(5)$ unification.
    Infinite sums, analogous to
Eqs. \eqref{eq:ground1}, \eqref{eq:basis1} were used to construct the modulo 5 basis from $U(1)$ eigenstates, with a ground
state obeying the cluster property up to a  constant factor.  In this Appendix we show how to construct a modulo 5 basis using finite
sums, starting from a modulo 15 invariant basis.

We actually deal with a more general case, by starting from a basis $|k\rangle_N$ with modulo $N$ invariance (up to a phase) and
constructing from it a new basis $|k\rangle_p$ with a modulo $p$ invariance.  Here the integer $p$ is a divisor of
$N$, so that
\begin{equation}\label{eq:factorized}
N=pr~~~,
\end{equation}
 with $r$ also an integer.

We start from the basis $|k,\omega\rangle_N$ of Eq. \eqref{eq:basis1}, and since the value of $\omega$ is fixed in this discussion
we define
\begin{equation}\label{eq:newdef}
|k \rangle_N \equiv e^{-ik\omega}|k,\omega \rangle_N~~~,
\end{equation}
so that
\begin{equation}\label{eq:phasedefn}
|k+Ns\rangle_N=e^{-iNs\omega}|k\rangle_N ~~~.
\end{equation}
We do not require states to be unit normalized, and so can write clustering for the basis $|k\rangle_N$  in the form
\begin{equation}\label{eq:clustern}
|k_A+k_B\rangle_N= |k_A \rangle_N  |k_B \rangle_N
 \end{equation}
for widely separated subsystems $A,\, B$.
We will now show how to construct from the basis $|k\rangle_N$ a new basis $|k\rangle_p$ obeying
\begin{equation}\label{eq:phasedefp}
|k+ps\rangle_p= |k\rangle_p~~~,
\end{equation}
and also obeying clustering.

Let us define $|k\rangle_p$ by the finite sum
\begin{equation} \label{eq:sumdef}
|k\rangle_p=r^{-1}\sum_{n=0}^{r-1} e^{i (k+ pn) \omega} |k+pn\rangle_N~~~,
\end{equation}
from which we have
\begin{equation}\label{eq:shifted}
|k+ps\rangle_p=r^{-1} \sum_{n=0}^{r-1} e^{i[k+ p(n+s)]\omega } |k+p(n+s)\rangle_N~~~.
\end{equation}
We can always decompose $n+s$ to exhibit its structure modulo $r$,
\begin{equation}\label{eq:modulor}
n+s=n^{\prime}+s^{\prime}r~~~,
\end{equation}
with $n^{\prime}$ and $s^{\prime}$ integers, and $0 \leq n^{\prime}\leq r-1$.
As $n$ ranges from $0$ to $r-1$, the residue $n^{\prime}$ also takes all
values in the range from $0$ through $r-1$, but in general in a different order.
Substituting Eq. \eqref{eq:modulor} into Eq. \eqref{eq:shifted} we have
\begin{equation}\label{eq:shifted1}
|k+ps\rangle_p=r^{-1} \sum_{n=0}^{r-1} e^{i(k+ pn^{\prime})\omega} e^{i  Ns^{\prime} \omega} |k+pn^{\prime} +N s^{\prime} \rangle_N~~~,
\end{equation}
where we have used $pr=N$  on the right hand side.  Using  Eq. \eqref{eq:phasedefn}, the state vector on the right
becomes
\begin{equation}\label{eq:reduction}
|k+pn^{\prime} +N s^{\prime} \rangle_N=e^{-i  N s^{\prime}\omega} |k+pn^{\prime}\rangle_N~~~,
\end{equation}
and so rewriting the sum over $n$ as a sum over $n^{\prime}$, Eq. \eqref{eq:shifted1} simplifies to
\begin{equation}\label{eq:shifted2}
|k+ps\rangle_p=r^{-1} \sum_{n^{\prime}=0}^{r-1} e^{i(k+ pn^{\prime})\omega} |k+pn^{\prime}\rangle_N=|k\rangle_p~~~,
\end{equation}
as needed.

To show clustering, we rewrite Eq. \eqref{eq:phasedefp} as
\begin{equation}\label{eq:phasedefp1}
|k\rangle_p=|k+ps\rangle_p ~~~,
\end{equation}
and average over $s$ to get
\begin{equation}\label{eq:phasedefp2}
|k\rangle_p=r^{-1}\sum_{s=0}^{r-1}|k+ps\rangle_p ~~~.
\end{equation}
Using Eq. \eqref{eq:sumdef} applied to the state $|k+ps\rangle_p$,
this can be written as the double sum
\begin{equation}\label{eq:doublesum}
|k\rangle_p=r^{-2}\sum_{s=0}^{r-1}\sum_{n=0}^{r-1}  e^{i[k+ p(s+n)]\omega}|k+p(s+n)\rangle_N~~~.
\end{equation}
When $k=k_A+k_B$, associating $s$ with the subsystem $A$ and $n$ with the subsystem $B$, and assuming that for widely separated $A$ and $B$  the state $|k+p(s+n)\rangle_N$ becomes the tensor product
\begin{equation}\label{eq:tensorprod}
|k_A+ps+k_B+pn\rangle_N=|k_A+ps\rangle_N |k_B+pn\rangle_N~~~,
\end{equation}
Eq. \eqref{eq:doublesum} reduces to
\begin{align}\label{eq:doublesum1}
|k_A+k_B\rangle_p=&r^{-2}\sum_{s=0}^{r-1}\sum_{n=0}^{r-1} e^{i[k_A+k_B+ p(s+n)]\omega}|k_A+ps\rangle_N |k_B+pn\rangle_N\cr
=&r^{-1}\sum_{s=0}^{r-1}  e^{i(k_A+ps)\omega}|k_A+ps\rangle_N
~~\times~~  r^{-1}\sum_{n=0}^{r-1}  e^{i(k_B+ pn)\omega} |k_B+pn\rangle_N\cr
=&|k_A\rangle_p  |k_B\rangle_p~~~.\cr
\end{align}
Thus the modulo $p$ state basis $|k\rangle_p$ inherits the clustering property of the modulo $N$
state basis $|k\rangle_N$.  Note that the norm squared of the modulo $p$ basis is related to the
norm squared of the modulo $N$ basis by
\begin{equation}\label{eq:normfinal}
{}_p\langle k|k\rangle_p=r^{-2} \sum_{n=0}^{r-1}{} _N\langle k+pn|k+pn\rangle_N~~~,
\end{equation}
so unit normalization of the modulo $N$ basis would not imply unit normalization of the
modulo $p$ basis.

\end{document}